# Homoepitaxy of rhombohedral-stacked MoS$_2$ with room temperature switchable ferroelectricity

Tilo H. Yang[1], Hsiang-Chi Hu[1], Fu-Xiang Rikudo Chen[2], Po-Yen Lin[3], Yu-Fan Chiang[1], Wen-Hao Chang[1], Yi-Hao Kuo[2], Yu-Seng Ku[4], Bor-Wei Liang[5], Alice Chinghsuan Chang[6], Han-Chieh Lo[1], Yu-Chen Chang[1], Yi-Cheng Chen[1], Ting-Hua Lu[1], Chun-Liang Lin[2]\*, and Yann-Wen Lan[1,7]\*

[1]Department of Physics, National Taiwan Normal University, Taipei 11677, Taiwan
[2]Department of Electrophysics, National Yang Ming Chiao Tung University, Hsinchu 30010, Taiwan
[3]Institute of Cellular and Organismic Biology, Academia Sinica, Taipei 11529, Taiwan
[4]Department of Physics, National Taiwan University, Taipei 10617, Taiwan
[5]Graduate Institute of Electronics Engineering, National Taiwan University, Taipei 10617, Taiwan
[6]Center for Measurement Standards, Industrial Technology Research Institute (ITRI), Hsinchu 30011, Taiwan
[7]Advanced Materials and Green Energy Research Center, National Taiwan Normal University, Taipei 11677, Taiwan

\*Corresponding authors. Email: clin@nctu.edu.tw (C.L.L.), ywlan@ntnu.edu.tw. (Y.W.L.)

**Abstract**

The discovery of interfacial ferroelectricity in two-dimensional rhombohedral (3R)-stacked semiconductors opens up a new pathway for achieving ultrathin computing-in-memory devices. However, exploring ferroelectricity switching in natural 3R crystals is difficult due to lack of co-existing 3R stacking domains. Here, we present that $MoS_2$ homoepitaxial patterns with 3R polytypic domains can manifest switchable ferroelectricity at room-temperature. Based on the diffusion limited aggregation theory, such $MoS_2$ patterns are formed under the low Mo chemical potential and low temperature with respect to common chemical vapor deposition synthesis. The alternation of 3R polytypes in the $MoS_2$ homoepitaxial patterns, observed by scanning transmission electron microscopy, accounts for ferroelectricity switching. The $MoS_2$ field-effect transistors with 3R polytypic domains exhibit a repeatable counterclockwise hysteresis with gate voltage sweeping, an indication of ferroelectricity switching, and the memory window exceeds those measured for compact-shaped 3R bilayer devices. This work provides a direct growth concept for layered 3R-based ferroelectric memory.

Ferroelectric (FE) materials, generally featuring polar point group in their unit cell, play a vital role in many electronic applications like non-volatile memories and pyroelectric sensors[1-6], to name a few. In particular, ferroelectricity combined with small band gap and high mobility semiconductors enables the new function of computing-in-memory[5,7]. Meanwhile, thinning down FEs is an inevitable trend for device miniaturization and lower power consumption. However, the FE dipoles perpendicular to the film surface are difficult to be retained due to the depolarization effect when conventional FEs reach the atomic-thickness regime[2,7,8]. To overcome this issue, theoretical studies have predicted that ferroelectricity can be manifested in van der Waals (vdW) interfaces between two marginally twisted 2D materials[9,10]. The clean vdW interface is able to maintain the FE property intact thanks to the absence of dangling bonds. This intriguing phenomenon has recently been experimentally demonstrated in 2D materials that lack polar point groups in their parent lattices, including boron nitride (BN) and transition metal dichalcogenides (TMDs)[11-14]. To produce the interlayer charge transfer, stacking sequence in 2D vdw FEs should exclusively be rhombohedral (3R) stacking. For memory applications, the capability of FE polarization flipping is required. The marginally stacked 3R TMDs have shown the capability of FE polarization switching by applying a vertical electric field, emphasizing the importance of the pre-existing oppositely-polarized 3R domains because they reduce the energy barrier for polarization switching[12,13]. Natural 3R-stacked TMDs grown under equilibrium tend to form single compact triangular and hexagonal crystal with perfect 2H or 3R stacking, thereby losing its capability of polarization switching. To this end, a synthesis approach to directly grow 2D layered semiconductor with well-organized 3R domains is highly desirable.

As the counterpart, a non-equilibrium growth regime often leads to dendritic/fractal growth of 2D TMDs[15-18]. According to the diffusion-limited aggregation (DLA) model, dendritic growth is predominantly attributed to the relatively fast diffusion of adatoms on a free surface with respect to the diffusion along an island edge (edge relaxation)[19,20]. Diffusion process in this model needs adatoms to hop among energetically favorable sites on the substrate and along the edge. The resultant crystals usually feature preferred growth direction following the crystal orientation of the substrate, analogous to the epitaxial behavior. In 2D materials synthesis, several process factors have been proven to trigger anisotropic growth, including substrate symmetry[15], environmental chemical potentials[18], and precursor mediation[21]. Recent works have shown that large scale, highly organized TMDs homoepitaxial patterns were realized via chemical vapor deposition (CVD) under certain conditions[15,18], suggesting that the precise control over dendritic growth is a promising approach to tailor stacking type, edge configuration, and domain arrangement in 2D vdW materials epitaxy. As can be expected, manipulation of stacking sequences in 3R dendrites enables ferroelectricity switching, creating a direct, high-yield growth method for 2D vdW FEs with respect to the artificial stacking counterpart.

In this work, we demonstrate the successful synthesis of 3R-stacked $MoS_2$ bilayer with room-temperature switchable ferroelectricity through a deliberately controlled CVD process. By tuning the growth under a low temperature, sulfur-rich condition, the subsequent $MoS_2$ crystal growing on the underlying $MoS_2$ surface has high anisotropic tendency in the initial stage, forming multiple nanoribbons aligned along preferred growth directions. Polarized spectroscopic analysis confirms the high degree of circular polarization for 3R-stacked $MoS_2$. By conducting a scanning transmission electron microscopy (STEM), the anisotropically grown $MoS_2$ bilayer is identified to mainly be 3R-like stacking while 3R domains are alternately embedded. With combining scanning tunneling spectroscopy (STS) and field-effect transistor (FET) characterizations, we robustly detect room-temperature switchable ferroelectricity in the anisotropically branched 3R $MoS_2$, in contrast to compact-shaped 3R $MoS_2$ that barely has the ability of polarization switching. Along with the proposed growth mechanism based on the diffusion limited aggregation (DLA) theory, 3R-stacked $MoS_2$ with various crystal sizes, layer numbers and shapes can be reproducibly obtained and reliably grown on both $SiO_2$/Si and sapphire substrates.

**Results and Discussion**

Figure 1a illustrates three representative polytypes in 3R stacked bilayer MoS$_2$, namely, AA, AB, and BA stackings. They are distinct in atom packing orders between layers. The same atoms from upper and lower layers are overlapped along the z-axis in the AA stacking, whereas the S atoms from the upper layer are located on the top of the Mo atoms from the lower layer in the AB stacking, and vice versa in the BA stacking. Since different 3R stacking polytypes correspond to the opposite FE polarizations in the out-of-plane direction, they are of interest and investigated in this work.

To achieve 3R stacked polytypes in single bilayer flake, the homoepitaxial growth of bilayer MoS$_2$ via a modified CVD synthesis is conducted as depicted in the Experimental Section and is shown in Fig. 1b. In brief, a two-step CVD synthesis was performed at 800 °C for 3 min and 670 °C for 17 min, respectively. The higher temperature selected for the first step was to intensively vaporize MoO$_3$ precursor subsequently brought to the growth substrate by Ar carrier gas. During the second step, the S zone was maintained at 250 °C, ensuring sufficient S supply for the nucleation and growth of MoS$_2$. Remarkably, a great number of MoS$_2$ bilayers with threefold symmetry, similar in appearance to fan, were emerged on SiO$_2$/Si substrate (Fig. 1c). Such a 3-fold symmetrical structure is unusual compared to most of reported bilayer MoS$_2$ featuring either homogeneous growth or random distribution on the underlying monolayer[22-24]. To clarify the growth environment responsible for fan-shaped MoS$_2$, duration time of the Mo zone was tuned, as shown in Fig. 1d. When the first duration at 800 °C was decreased to 2 min, only monolayer MoS$_2$ crystals were produced, indicating that the Mo amount is not sufficient to promote the vertical growth. On the other hand, one-step process was separately conducted at 800 °C and 670 °C for 20 min. Bilayer MoS$_2$ with compact triangular/hexagonal shapes was observed in the former case, while MoS$_2$ nucleation barely took place on the entire substrate in the latter case. These results convincingly suggest that the Mo supply for MoS$_2$ growth was mainly determined by the first duration at 800 °C. However, growth under the continuous Mo supply at high temperature led to regular compact-shaped bilayer MoS$_2$. A confined Mo amount and relatively low temperature is a key condition for 3-fold bilayer MoS$_2$ growth.

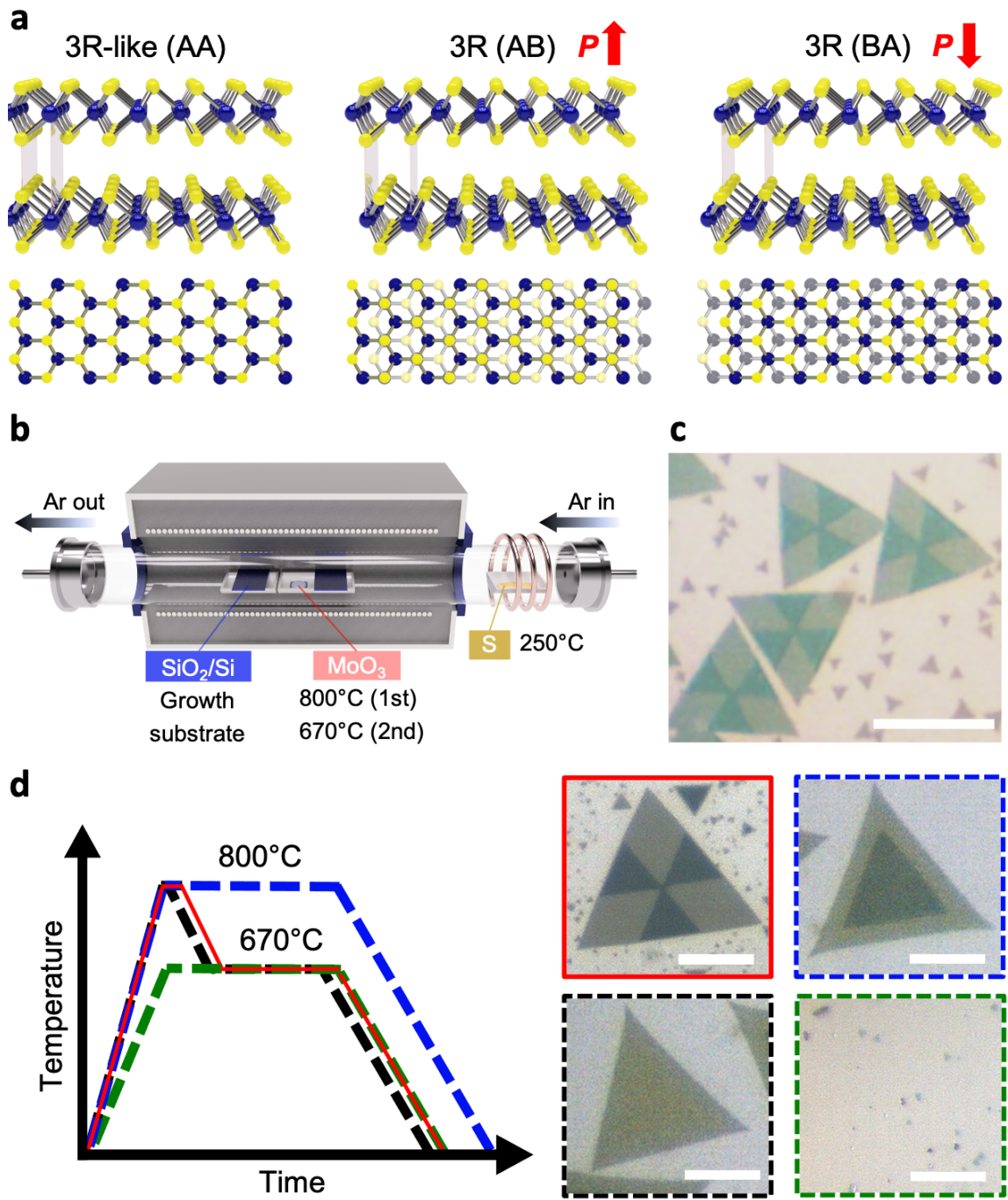

**Fig. 1 | a,** Schematics of the AA, AB, and BA stacking sequences in 3R bilayer MoS$_2$. Bottom layers are represented by lighter atoms. Mo: blue spheres; S: yellow spheres. The AB and BA stacking configurations possess the up and down interlayer FE polarizations, respectively. **b,** CVD setup for the homoepitaxial growth of MoS$_2$ with 3-fold symmetry. **c,** Optical image showing 3-fold homoepitaxial MoS$_2$ patterns on SiO$_2$/Si. Scale bar, 20 μm. **d,** Modulation of growth temperature and the resultant MoS$_2$ crystals. Scale bars, 10 μm.

The observed growth behavior in fan-shaped MoS$_2$ is originated from anisotropic growth rates of different edges. In thermodynamics, the growth rate of a crystalline edge can be quantitatively determined by the relative edge stability, characterized by the formation energy ($\gamma$) for each edge. Given that the edge formation energy is varied with the chemical potential of sulfur ($\mu_S$), one can elucidate the final equilibrium shape of MoS$_2$ under different $\mu_S$ conditions by means of the Wulff construction rules[25]. Figure 2a demonstrates the polar plot of the formation energy for the arbitrary edges under the S-rich environment. Notably, the armchair edges with a period of 60° exhibits the highest $\gamma$ value, suggesting that the six crystallographically-equivalent armchair directions are the most unstable and should own the relatively high growth rate during the CVD synthesis. The fast-growing edges eventually vanish and the final crystal will be faceted with the slow-growing edges. In such scenario, MoS$_2$ crystals will end up with a compact triangular shape with the zigzag edges as denoted by the red triangle in Fig. 2a, which evidently does not match the observed 3-fold homoepitaxial growth behavior. There were other factors involved in the growth; however, the large free energy of the armchair directions under the high $\mu_S$ provides the possible driving force during the MoS$_2$ bilayer growth.

Due to deviation from the equilibrium Wulff shape, the fan-shaped bilayer MoS$_2$ growth is intuitively linked with the non-equilibrium factors. Unlike the first MoS$_2$ layer growing on amorphous SiO$_2$/Si substrate, the subsequent MoS$_2$ layer was formed on the crystalline MoS$_2$ layer. In this regard, it should consider preferential diffusion pathways for precursor adatoms migrating on MoS$_2$ surface and competition with edge diffusion when adatoms attached to the second layer island. Fig. 2b illustrates the interplay between surface and edge diffusion of precursor monomer on 1H-MoS$_2$ surface, where amorphous MoS$_3$ molecule is considered as the main species during growth according to the Mo-S phase diagram[26]. Based on recent in-situ TEM observations and DFT calculation[27], we herein assume that the most energetically favorable adsorption site for molecular MoS$_3$ locates on the top of Mo atoms (denoted as T$_{Mo}$). The preferred kinetic pathway among these Mo sites is along the T$_{Mo}$-H-T$_{Mo}$ direction (the equivalent $\langle \bar{1}2\bar{1}0 \rangle$ direction, as denoted by red arrows in Fig. 2b), where H denotes the center of the hexagon or the hollow site. When MoS$_3$ monomer reaches a MoS$_2$ bilayer terrace, an additional stabilization energy is given by the terrace lattice due to the in-plane bonding. According to the DLA theory[19,20], such diffusion dynamics can be expressed to be

$$\Gamma_s = v e^{-E_{D,s}/k_B T} \quad (1)$$

$$\Gamma_e = v e^{-(E_{D,s}+\Delta E_{stab})/k_B T} \quad (2)$$

, where $\Gamma_s$ and $\Gamma_e$ is the hopping rate of a monomer among two neighbor adsorption sites on a surface and along an island edge, respectively, $v$ is the vibrational frequency of a monomer, $E_{D,s}$ is the potential energy barrier to surface diffusion, $\Delta E_{stab}$ is the stabilization energy by the edge, $k_B$ denotes the Boltzmann constant, and $T$ is the substrate temperature. For MoS$_2$ homoepitaxial growth, the $E_{D,s}$ is similar in surface diffusion and edge diffusion since the MoS$_3$ molecules migrate along the energetically favorable pathways in the both scenarios. Given that an additional energy $\Delta E_{stab}$ is contributed to the energy barrier for edge diffusion, the relative dominance between the two diffusion regimes is closely correlated with the temperature. The competition between surface and edge diffusion is numerically simulated using their Arrhenius relation, as plotted in Fig. 2c. The degree of local relaxation along the island edge was affected by the ratio of the hopping rates, $R_D=\Gamma_s/\Gamma_e$, which significantly varies with temperature. At higher temperature, compact, high uniform bilayer island could be formed with the sufficient edge diffusion mobility. However, at low temperature the ability of edge diffusion for adatoms is limited, leading to the fractal/dendritic growth. The growth direction of those dendrites reflects the preferred pathways oriented by the lattice geometry of the underlying layer. Fig. 2d shows the morphological evolutions of fan-shaped bilayer MoS$_2$ with increasing the 2$^{nd}$ duration time at 670 °C in which $R_D$ is ≈$10^7$, for detailed synthesizing parameters are depicted in Supplementary Fig. 1. The fan-shaped bilayer growth started from self-aligned nanoribbon-like structures, according well with the predicted growth behavior in the DLA model. Investigation of atomic force microscopy (AFM) confirms that these ribbons were bilayer with high uniformity in thickness (Fig. 2e). Depending on the order of growth, the primary and secondary bilayer dendrites are readily distinguished. The bilayer dendrites were all aligned along the preferential kinetic pathways defined by the underlying MoS$_2$ layer's lattice orientation and the angle between them is approximately 60°. Such preferred migration directions have the equivalent vectors in the $\langle \bar{1}2\bar{1}0 \rangle$ directions which is the same with the armchair directions. This explains the origin why those bilayer dendrites tended to be arranged parallel to the edge the first layer. These results supports the fact that the fan-shaped bilayer MoS$_2$ growth belongs to the epitaxial growth regime. Importantly, different fractal-like regimes, characterized in terms of the branch width and numbers of initial dendritic bilayer, could be modulated with different $R_D$ by changing the 2$^{nd}$ duration temperature (see Supplementary Discussion 1 and Supplementary Fig. 2). Adjacent parallel bilayer branches were finally impinged with each other to coalesce into a complete single-crystalline bilayer triangle, forming 3-fold homoepitaxial patterns.

After establishing the growth model, we examine the versatility of this model to other substrates and geometry of the first layer, as demonstrated in Supplementary Fig. 3. Fan-shaped bilayer MoS$_2$ crystals were successfully grown on c-plane sapphire via our modified CVD process. The grown MoS$_2$ on sapphire also exhibited the similar material characteristics as in the SiO$_2$ case. This further proves that the 3-fold bilayer growth was not guided by lattice symmetry of growth substrates. Additionally, the fan-MoS$_2$ crystals grown on SiO$_2$ were tuned to be hexagonal shape, implying that the geometry of the parent monolayer had little influence on the 2$^{nd}$ layer growth.

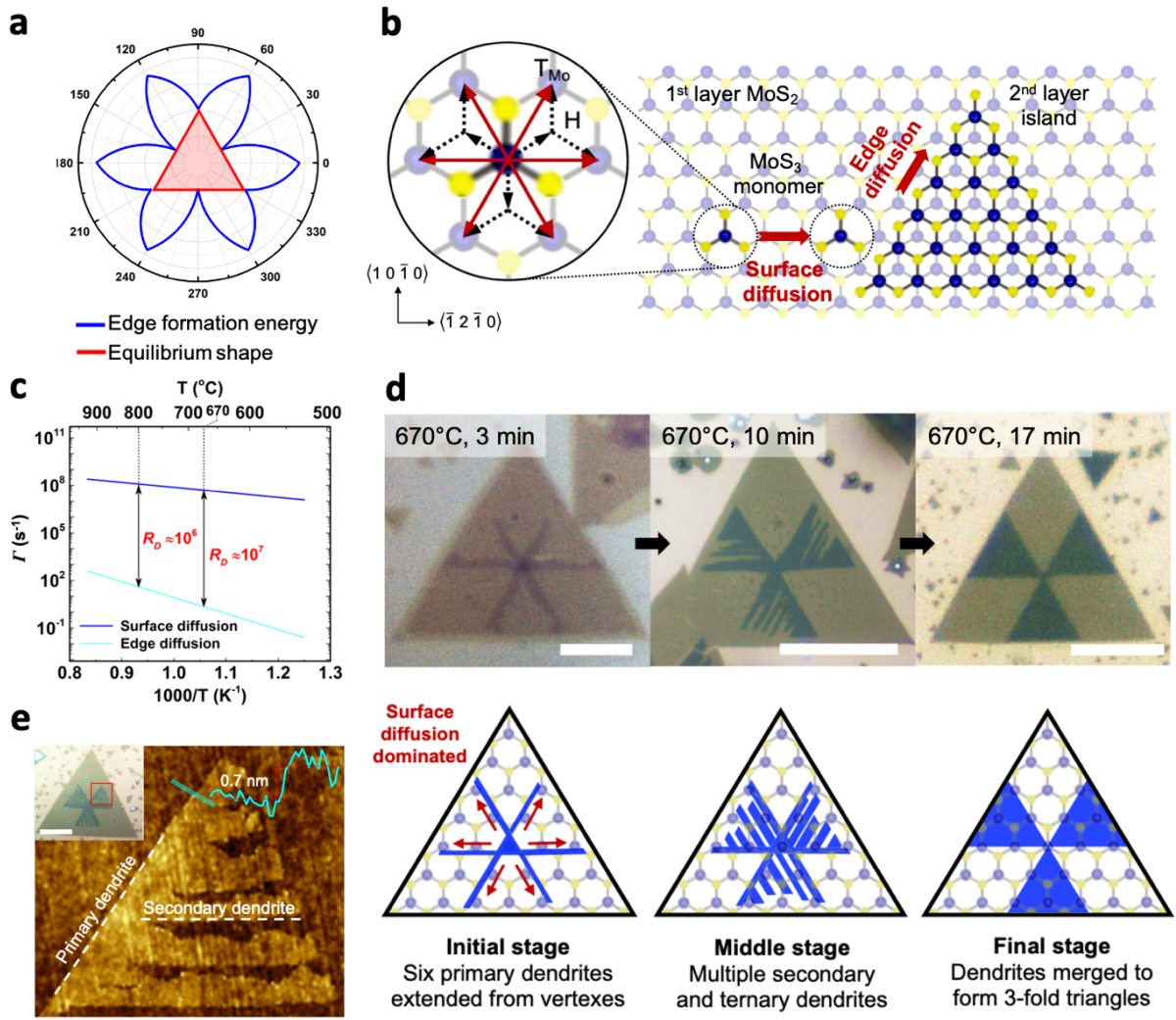

**Fig. 2 | Proposed growth mechanism of 3-fold MoS₂ homoepitaxy. a,** Polar plots of the formation energies of MoS₂ edges as a function of chiral angle χ (0° ≤ χ ≤ 360°) under a S-rich environment. The interior red triangle represents the final equilibrium shape of MoS₂ flake. The 0° corresponds to the armchair direction. **b,** Schematic of surface and edge diffusion for MoS₃ monomer on 1H-MoS₂ surface. **c,** Arrhenius relation of the hopping rates of MoS₃ monomer for surface and edge diffusion. **d,** Optical images and the corresponding schematics demonstrating the morphological evolutions of fan-shaped bilayer MoS₂ with duration time of the Mo zone. Scale bars, 20 μm. **e,** AFM image showing the nanoribbon structure of the fan-shaped MoS₂, obtained from the red rectangle in the inset optical image. The primary and secondary dendrites are clearly observed. The thickness profile is taken along the green marker. Scale bar, 10 μm.

To inspect atomic structures in fan-shaped MoS₂ homoepitaxial crystals, transmission electron microscope (TEM) is utilized. The selective area diffraction patterns (SADP) on bilayer and monolayer regions are respectively shown in Figure 3a,b. Both regions present one set of diffraction spots arranged in the same rotation angle, meaning that they were single crystals oriented in the crystallographically-equivalent direction (0° or 60°). Information of stacking configurations can be obtained from the relative intensity profile extracted from those diffraction spots. The $\{1\,0\,\bar{1}\,0\}$ spots were evidently weaker than the $\{1\,1\,\bar{2}\,0\}$ spots (left panels in Fig. 3a,b) in the bilayer's SADP, according well with the simulated SADP results for the 3R phase[28]. The 3R stacking results are also consistent with the optical microscopic image where fan-shaped bilayers rotate 0° with respect to the underlying $MoS_2$ monolayer (Fig. 1c). Complete information on the atom packing orders in fan-shaped bilayer was acquired by performing the scanning TEM-annular dark field (STEM-ADF) imaging. Fig. 3c demonstrates a representative ADF image taken from the edge between monolayer and bilayer nanoribbon as indicated in the inset. The Mo and S atoms can be readily distinguished from the ADF contrast since the contrast is proportional to the atomic number (Mo(42) and S(16))[29,30], and heavier atoms exhibit brighter contrast. Intriguingly, such a local bilayer nanoribbon possesses 3R (AB and BA) and 3R-like (AA) domains alternately, with interlayer sliding (partial dislocation) between them. Statistically, a larger percentage of bilayer area was predominantly occupied by the AA stacking (Fig. 3d), while AB and BA domains are distributed between the AA-stacked domains (Fig. 3e). Fig. 3f-i further display the atomic arrangement of the AB and BA domains at a bilayer/monolayer edge. Since the AB and BA domains hold mirror symmetry and demonstrate similar ADF contrasts, it is difficult to identify them when ADF imaging was taken from the domain center. The evolution of intensity profile from monolayer to bilayer is important to identify the AB/BA stacking. Fig. 3g,i show the intensity profiles along the periodic Mo sites from the monolayer to the bilayer region as indicated in Fig. 3f,h, respectively. Fig. 3g exhibits an abrupt increase from the monolayer to the bilayer, indicating the domain shown in Fig. 3f is the AB stacking since S atoms from the upper $MoS_2$ layer sit on the top of the Mo atoms from the lower $MoS_2$ layer. By contrast, Fig. 3i show almost the same contrast, meaning the domain shown in Fig. 3h is instead the BA stacking.

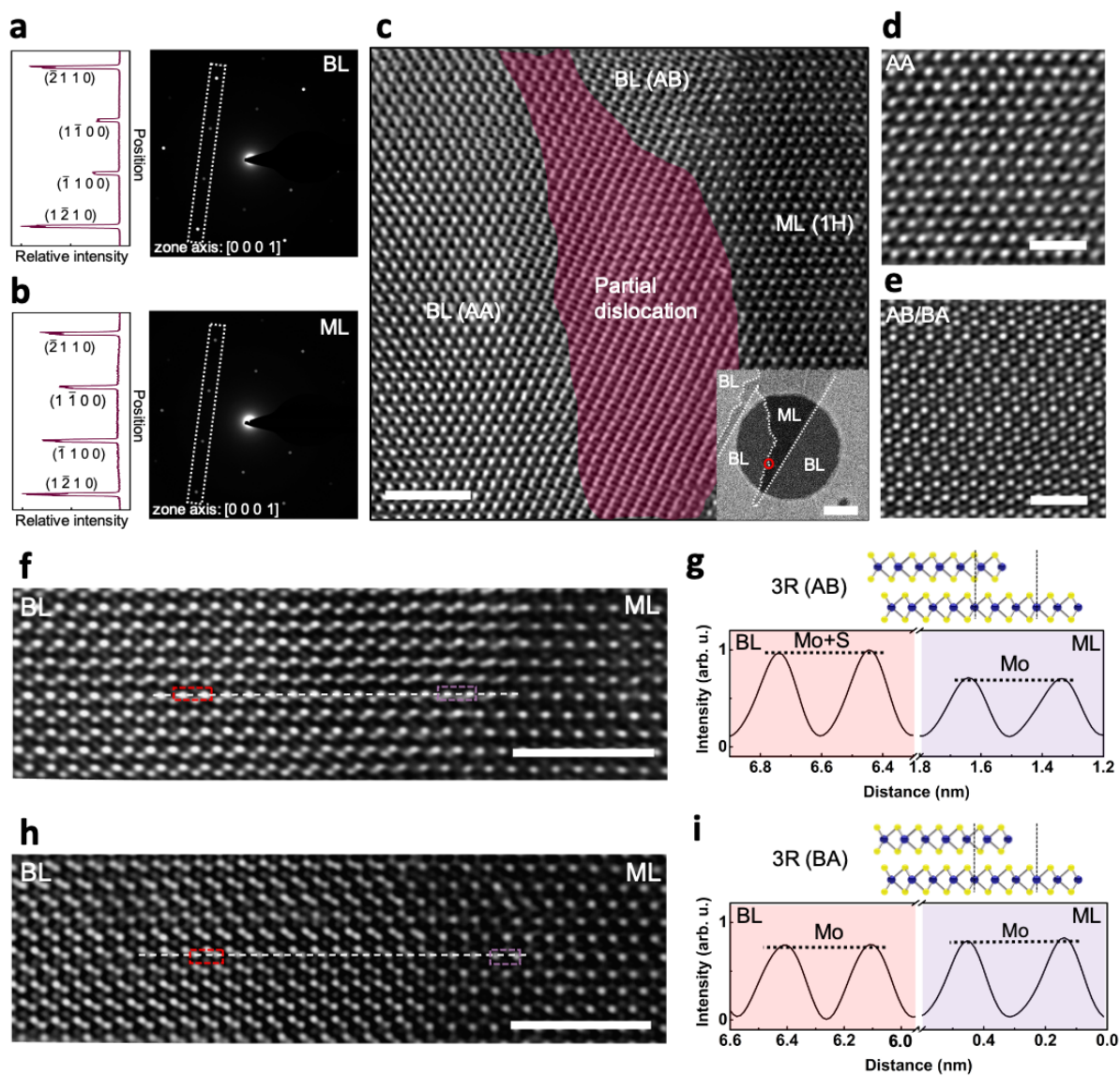

**Fig. 3 | a,b,** SADPs obtained from fan-bilayer and monolayer regions, respectively. The intensity profile was extracted along the white dash marker in the SADP. **c,** STEM-ADF image demonstrating the representative monolayer/bilayer nanoribbon edge, obtained from the red hollow circle in the inset. 3R and 3R-like domains are found separated by partial dislocation zone (stacking boundary). Scale bars, 2 nm in (**c**) and 200 nm in inset. **d,e,** STEM-ADF images showing the atomic arrangement of 3R and 3R-like domains, both of which were taken from the same bilayer ribbon. Scale bars, 1 nm. **f,h,** STEM-ADF images demonstrating the atomic configurations of the AB and BA stackings near fan-bilayer/monolayer edge. Scale bars, 2 nm. **g,i,** Intensity profiles extracted along the yellow lines across the monolayer/bilayer edge in (**f**) and (**h**).

To acquire basic material characteristics of fan-shaped $MoS_2$ bilayer, we systematically performed spectroscopic investigations, as shown in Figure 4. In photoluminescence (PL) analysis under a 532 nm laser as the excitation source (Fig. 4a), the intensity of the resonant peak from the monolayer was higher than that from the bilayer. However, we noticed that the fan-shaped bilayer flakes exhibit an anomalous blueshift in the A-exciton peak with respect to the monolayer. The blueshift was found common on the entire bilayer flake, as shown in the A-exciton peak position mapping of Fig. 4b. Interestingly, such phenomenon was not seen in a regular compact-shaped 3R bilayer, for detailed comparisons in PL spectra between regular 3R and fan-shaped bilayers are presented in Supplementary Fig. 4. The blueshift was also observed in the fan-bilayer in the ribbon-like stage. Fig. 4c presents the spatial distribution of fan-shaped $MoS_2$ for Raman peaks separation (Δ) between the out-of-plane $A_{1g}$ (≈405 $cm^{-1}$) and in-plane $E^1_{2g}$ (≈385 $cm^{-1}$) vibration modes under 532 nm laser excitation. Though the difference in Raman peak separation for the monolayer (Δ=18.7 $cm^{-1}$) and the fan-bilayer (Δ=21.6 $cm^{-1}$) is rational due to the difference in thickness, the Δ from fan-bilayer was found smaller than regular bilayer's (Δ=23.3 $cm^{-1}$). Detailed Raman spectra obtained from the two distinct bilayers are provided in Supplementary Fig. 5. The $E^1_{2g}$ mode shows a noticeable blueshift (≈1.7 $cm^{-1}$) in the fan-shaped bilayer compared with the regular 3R bilayer. This, correlated with the observed blueshift in the PL A-exciton peak, can be attributed to residual compressive strain in 2D TMDs[31]. The strain effect was induced by interlayer sliding among multiple 3R domains, supported by the STEM investigations. To rule out the strain effect resulting from defect-induced variation of elemental compositions[32], X-ray photoelectron spectroscopy (XPS) was performed. The Mo/S ratios from the monolayer and bilayer regions are close to 1:2, and the elemental characteristic peaks (i.e., Mo 3d and S 2p) do not exhibit noticeable chemical shift (Supplementary Fig. 6), revealing that both regions are stoichiometrically equivalent.

To further verify the origin of the observed blueshift in PL, mapping the electronic band structure of fan-shaped $MoS_2$ crystals is necessary. For this, a high-resolution scanning tunneling microscopy/spectroscopy (STM/S) is conducted. Numerous fan-shaped $MoS_2$ samples were transferred to a highly oriented pyrolytic graphite (HOPG) substrate for STM measurement. Fig. 4d shows selective *dI/dV* spectra from monolayer and bilayer in an atomic-resolution STM image of Fig. 4e. The detailed evolution of band structure across the monolayer/bilayer edge is shown in Supplementary Fig. 7. In a *dI/dV* spectrum, the position of the valence band maximum (VBM) and the conduction band minimum (CBM) can be estimated, leading to the bandgap ($E_g$) = $E_{CBM}$ - $E_{VBM}$. The $E_g$ of the monolayer and bilayer was measured to be 1.9 eV and 1.7 eV, respectively. This further confirms that the observed blueshift in PL spectra was not caused by the larger bandgap in fan-bilayer flake but by local strain instead. Furthermore, the real-space STM image and the corresponding Fourier transform of both regions clearly shows the same orientation (see inset of Fig. 4e), indicating that the bilayer flake shares the same crystal orientation with the monolayer. These results support the fact that fan-shaped bilayer $MoS_2$ belongs to an epitaxial growth mechanism.

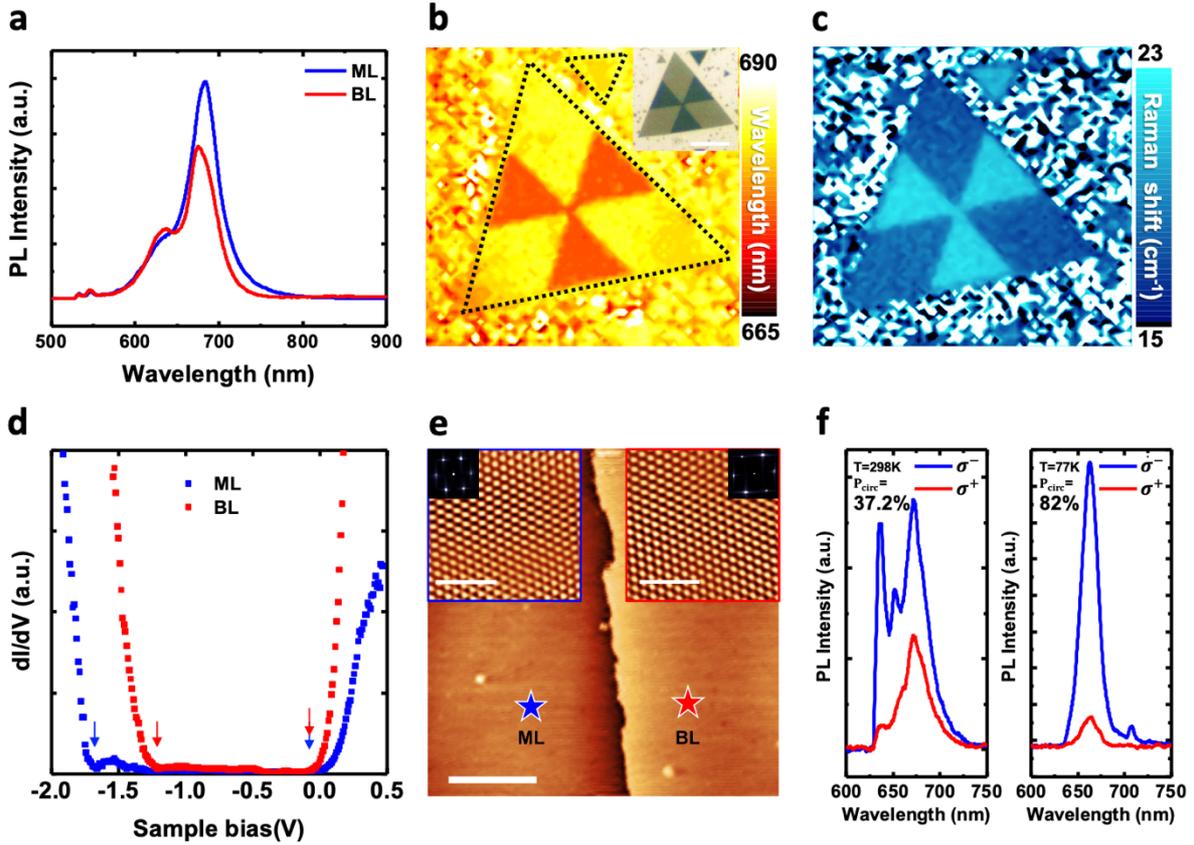

**Fig. 4 | Spectroscopic characterizations of fan-shaped MoS₂. a,** Representative PL spectra taken from monolayer and bilayer domains. **b,** Spatial PL mapping for the A-exciton peak's position taken from the inset OM image. Scale bar, 10 μm. **c,** Spatial Raman mapping for the peak separation between the $E^1_{2g}$ and the $A_{1g}$ modes. **d,** A selective subset of $dI/dV$ spectra acquired on the monolayer and bilayer surface as marked in (**e**). The arrows indicate the edges of the valence and conduction bands. **e,** Typical STM topography image taken near the monolayer/bilayer edge. The insets show the atomic-resolution structure and the corresponding Fourier transform of the monolayer (left) and bilayer (right), respectively. Scale bars, 50 nm in (**e**) and 2 nm in inset. **f,** Representative $\sigma^-$ (blue) and $\sigma^+$ (red) PL spectra obtained under $\sigma^-$ 633 nm excitation at 298 K (left) and 77 K (right) on the bilayer domain. The measured DOP was 37.2% and 82%, respectively.

Circular dichroism in stacked TMDs is a stacking-dependent physical property[23,33,34], thereby being a solid approach for examining stacking configuration. Given that I($\sigma^-$) and I($\sigma^+$) denote the polarization-resolved PL intensity under left- and right-handed polarized excitation, the degree of circular polarization ($P_{circ}$) can be defined as

$$P_{circ} = \frac{I(\sigma^-) - I(\sigma^+)}{I(\sigma^-) + I(\sigma^+)} \quad (3)$$

Fig. 4f shows the $P_{circ}$ of fan-shaped bilayer MoS₂ crystal at room temperature (~298 K) and low temperature (~77 K, liquid nitrogen) under the excitation of 633 nm continuous-wave laser source. The bilayer flake exhibits the higher $P_{circ}$ of ~38% compared with the 20% for monolayer under ambient conditions, and shows even higher $P_{circ}$ of ~82% at 77 K. The spatially resolved $P_{circ}$ mapping are summarized in Supplementary Fig. 8, where all the results point out remarkably high $P_{circ}$ from the fan-shaped bilayer. Unlike the 2H phase which preserves inversion symmetry, the 3R stacking exhibits strong degree of polarization which is both spin- and valley-selective thanks to the broken inversion symmetry[34]. Consequently, the observed high $P_{circ}$ value in the fan-shaped MoS$_2$ bilayers is believed to reflect the 3R stacking configuration.

Based on the alternating 3R stacking domains in fan-MoS$_2$ bilayers, the ferroelectric characteristic is thoroughly studied. Figure 5a displays the tilted dark-field (DF) TEM image taken near the monolayer/bilayer edge, obtained by filtering the $(1\,0\,\bar{1}\,0)$ diffraction spot denoted in the inset of Fig. 5a. Considering that the DF-TEM image contrast for circling the $\{1\,0\,\bar{1}\,0\}$ (first order) family can provide the information of stacking features and the spot intensity varies periodically with the tilt angle[28,35,36], the contrast shown in Fig. 5a,b exhibits the alternation of AA, AB, and BA domains in a few microns range. Such co-existence of multi 3R domains in a large-scale is convenient not only for examination of ferroelectric switching by the fabricated fan-shaped MoS$_2$ FET but also for future device applications. Supplementary Fig. 9 reveals bilayer nanoribbons featuring asymmetrical edge morphologies, where the rough side corresponds to the growth front of bilayer ribbons that was impinged with the smooth side of the adjacent nanoribbons. We notice that the AB and BA domains tended to nucleate at the concave corner along the growth front of a bilayer ribbon (see Fig. 5a). This is possibly because lattice reconstruction at the slow growing corner is allowed to reach lower edge-energy configurations, namely AB and BA stackings. Edge recrystallization in fractal dendrites was also proposed in a recently established 2D DLA model[37]. Fig. 5c illustrates the transition between three kinds of 3R domains via local interlayer strain, and the associated energy barrier profile for domain transition is presented in Fig. 5d[9]. Compared to the AB and BA stacking, the AA stacking has the higher stacking energy due to the Coulomb repulsion between the same atoms from different layers. The initial AA stacking region tends to transform into the AB and BA stacking configurations with the ground state energy via interlayer sliding[14,38].

Earlier research has suggested that the AB and BA stackings in 2D TMDs manifest the opposite interlayer FE polarizations in the out-of-plane direction[12,13]. We examine the FE effect on the modulation of band structure by the cycling STS measurement. As shown in Fig. 5e,f, the tunneling current between the forward and backward bias sweeping exhibits hysteresis phenomenon in the fan-bilayer region, while two overlapping tunneling currents were presented by the monolayer region. The measured hysteresis in the cycling STS results proved the polarization switching in ferroelectric materials[39,40]. Based on co-existing 3R domains in a single fan-shaped MoS$_2$ bilayer flake, a prototype ferroelectric semiconductor

FET (FeS-FET) is fabricated, as schematically illustrated in Fig. 5g. A 10 nm-thick hBN layer was placed below fan-shaped MoS$_2$ in order to mitigate the charge trapping effect from SiO$_2$/Si substrate. Fig. 5h shows the representative transfer curve of a fan-shaped MoS$_2$ FeS-FET. The device exhibits a counterclockwise hysteresis behavior, which is a typical signature of FE switching. By contrast, the regular 3R bilayer MoS$_2$ FET (Fig. 5i) reveals a much smaller counterclockwise hysteresis window, implying the limited ability of FE switching due to the absence of co-existence of 3R domains. As shown in Supplementary Fig. 10, the hysteresis phenomenon is repeatable and reproducible in other devices, and retention measurement was conducted to evaluate the endurance of FE polarization. For comparison, the monolayer MoS$_2$/hBN FET was fabricated and shows almost a hysteresis-free transfer curve (Supplementary Fig. 10). The FET characterizations proved that interfacial ferroelectricity in fan-shaped bilayer MoS$_2$ can be flipped by sweeping a back gate at room temperature.

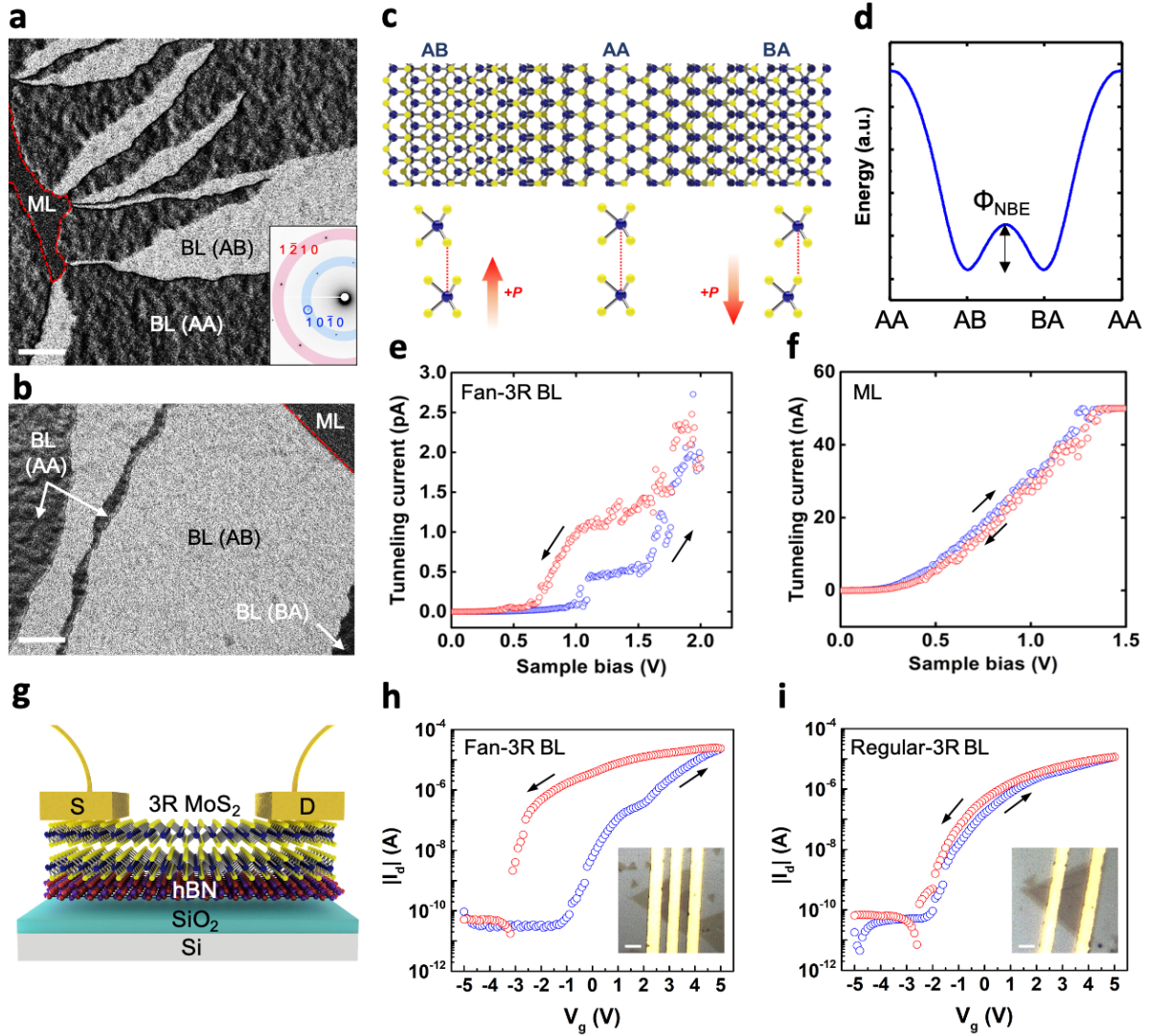

**Fig. 5 | Ferroelectricity switching in fan-shaped bilayer MoS₂. a,b,** Tilted DF-TEM images demonstrating alternating 3R/3R-like domains formed in bilayer ribbons. The red dash lines mark the edge of bilayer nanoribbons. Scale bars, 100 nm. **c,** Schematic of transformation between 3R-like (AA), 3R (AB), and 3R (BA) domains induced by local strain of the upper and lower layers. The AB and BA domains are responsible for the ferroelectricity with polarization up and down in the out-of-plane direction, respectively. **d,** The energy barrier in between 3R domains transformation. The tunning current in STS of (**d**) fan-bilayer and (**e**) monolayer regions. **f,** Schematic of 3R-stacked bilayer MoS₂ FET. Transfer characteristics of (**g**) fan-shaped and (**h**) regular-shaped 3R bilayer MoS₂ FET. Scale bars, 5 μm.

## Conclusion

In summary, we have successfully synthesized 3R bilayer MoS$_2$ with switchable interfacial ferroelectricity via two-step CVD method. Through suppressing the edge diffusion mobility with $R_D \approx 10^7$ and under the high sulfur chemical potential, bilayer MoS$_2$ crystals were tailored from compact shape with single 3R domain to 3-fold distributed nanoribbons with multiple 3R domains. We find that the AB and BA domains, corresponded to oppositely polarized FE domains, tended to nucleated in the initial AA domains, forming a hybrid 3R domains microstructure. The counterclockwise hysteresis in the fan-shaped 3R $MoS_2$ FET indicates that the ferroelectric polarization could be manipulated by applying a back gate. This work provides the engineering of stacking configurations in 2D TMD crystals via growth process. The direct growth of 3R FE semiconducting materials is beneficial to high yield manufacturing for computing-in-memory applications.

**Methods**

**CVD Synthesis.** The homoepitaxial growth of bilayer $MoS_2$ was achieved by using an ambient-pressure CVD system. Prior to synthesis, 90 nm $SiO_2$/Si substrates used for growth were ultrasonically cleaned sequentially in acetone for 10 min, isopropanol for 5 min, and deionized water for 10 min, and finally dried with nitrogen gas. High quality S (powder, 99.99% purity) and $MoO_3$ (powder, 99.99% purity), purchased from Sigma-Aldrich, were used as the precursor sources. The reaction furnace is a quartz tube (4-inch in diameter) installed with three heating coils. A schematic drawing of the CVD setup is portrayed in Fig. 1c. A ceramic boat containing 10 mg $MoO_3$ was placed at the center of the furnace tube. $SiO_2$ substrate for growth, the polished side facing downward, was put on another ceramic boat placed 0.5 cm behind the $MoO_3$ boat. A crucible containing 350 mg S powder was located upstream in excess. High purity Ar (≥99.99%) as the carrier gas was passed at 424 sccm flow rate during the entire growth process. Two-step reaction temperature profile was used; the furnace was first ramped up with a rate of 25.7 °C/min to 800 °C and held at 800 °C for 3 min, and was then cooled down to 670 °C in 5 min and stabilized for 17 min. The S zone started to heat 2 min before the $MoO_3$ zone reached 800 °C, and kept among 200-250 °C when the $MoO_3$ zone underwent the isothermal process at 670°C. The furnace underwent natural furnace cooling to room temperature.

**Wet transfer process.** Poly (methyl methacrylate) (PMMA) was used as the supporting film to peel off the $MoS_2$ crystals from $SiO_2$ substrate. PMMA was spin-coated on the top of $MoS_2$ samples at 4000 rpm for 90 s, followed by baking at 110 °C for 3 min. The edge of the PMMA film was scribed by a blade, and was immersed slowly into an 80°C 0.03 M KOH solution. The PMMA film carrying $MoS_2$ samples was able to be easily peeled off after soaked in the KOH solution less than 30 min. Before picked up onto the destination substrate, the floating PMMA/$MoS_2$ film was subsequently rinsed with deionized water many times. After baking at 110 °C for 1h, the PMMA film was finally removed with acetone.

**Optical and topographic characterizations.** The optical properties of the $MoS_2$ film was characterized using Raman and PL spectroscopic measurements. The Raman and PL spectra were taken with an integrated confocal optical microscope system with a spectrometer (Kymera 328i, Andor) using a 532 nm (2.33 eV) continuous-wave laser as the excitation sources. A 100× objective lens with a numerical aperture of 0.9, and a variable 150 lines/mm and 1200 lines/mm grating was employed to obtain a better signal-to-noise ratio. For the circularly polarized PL measurements, a HeNe 633 nm (1.96 eV) continuous-wave laser was use as the excitation source. Measurements were performed at room temperature and low temperature using a liquid-nitrogen system. To achieve circularly polarized excitation, a quarter-wave plate (ThorLabs, Inc.) was used and the state of circular polarization was confirmed at the sample location. The laser spot was focused on the $MoS_2$ sample at normal incidence using an objective lens with a numerical aperture of 0.75, which also collected the emitted PL in the reflected direction. When the PL signal was returned, the quarter-wave plate converted the circularly polarized light to linearly polarized light that was then passed the analyzer to detect the left- and right-handed polarized light. By doing so, we obtained the helicity of PL signal to calculate the degree of circularly polarization.

For topographic measurement, a Nanoview 1000 AFM system was employed operating in the tapping mode, by using tip model Olympus OMCL-AC160TS (tip stiffness = 26 N/m, frequency = 300 kHz). The measurements were conducted under ambient conditions.

**STM/STS Characterization.** Fan-shaped $MoS_2$ samples were transferred onto a highly oriented pyrolytic graphite (HOPG) substrate. Before measurement, the $MoS_2$/HOPG substrate was subjected to vacuum annealing at 200 °C for 12 h in an UHV chamber in order to remove surface impurities. Fan-shaped $MoS_2$ sample was found by referring to its appearance in OM image. STM measurements (Omicron RT STM) were conducted in an UHV chamber with a background vacuum and temperature kept at $5\times10^{-10}$ Torr and room temperature, respectively. A tungsten STM tip used for measurement was manufactured by a sodium hydroxide chemical etching method.

**TEM Investigations.** To prepare TEM specimens, fan-shaped $MoS_2$ crystals were transferred onto a TEM grid (Quantifoil Mo grid). TEM observations were conducted using dark-field and STEM modes for stacking analysis and for atomic structure analysis, respectively. All the observations were performed on an aberration corrected JEM-ARM200FTH with a cold-field emission gun and a spherical-aberration corrector (the JEOL DELTA-corrector), operating at 80 kV. Parts of SADP were captured in a FEI Tecnai G2 F20 operating at 200 kV. All the experiments were performed at room temperature.

**XPS Characterization.** µ-XPS measurement was performed using a scanning photoemission spectroscopy (SPEM) at the 09A1 beamline of National Synchrotron Radiation Research Center (NSRRC) in Taiwan. Many fan-shaped $MoS_2$ samples were transferred on an Au-coated

SiO$_2$/Si substrate. Prior to measurement, the sample was pre-baked at 180 °C in vacuum at ~10$^{-9}$ Torr for 12 h, and XPS measurement was performed at room temperature. The photon source was an U5-SGM undulator, and the photon energy was set to 400 eV. The incident X-ray beam size was ≈0.1 μm. The passing energy was set to 0.75 eV.

**Device Fabrication and Electrical Characterization.** Fan-shaped MoS$_2$ samples were transferred onto a pre-patterned 100 nm SiO$_2$/Si substrate using PMMA-assisted wet transfer method. Mask for electrode deposition was made by spin-coating with PMMA photoresist at 4000 rpm for 55 s, followed by baking at 180 °C for 3 min. Source/drain electrodes were patterned performing e-beam lithography in a scanning electron microscope (JEOL JSM-6500F), and the substrate was then developed in an IPA/MIBK solution. Contact metals consisting of Bi (20 nm) and Au (40 nm) were deposited by a custom-built thermal evaporator (JunSun Tech Co. Ltd., Taiwan) at a rate of 1 Å s$^{-1}$. Acetone was used to do a final liftoff to obtain back-gated MoS$_2$ FETs. After the liftoff, the device substrate was baked at 120 °C for 3 min. Keithley 2636B source meter was used to acquire the current-voltage characteristics.

**Acknowledgements**

This work was supported by the Ministry of Science and Technology (MOST) of Taiwan through grant 108-2112-M-033-006, MOST-111-2119-M-008-003, MOST 110-2634-F-009-027, 110-2112-M-A49 -013 -MY3, and MOST 110-2112-M-A49 -022 -MY2. This work was also financially supported by the "Center for the Semiconductor Technology Research" from The Featured Areas Research Center Program within the framework of the Higher Education Sprout Project by the Ministry of Education (MOE) in Taiwan. The authors thank Ms. Yin-Mei Chang in The Instrumentation Center at National Tsing Hua University for her skilled TEM operation. The authors are grateful for technical support on SPEM data acquisition from Mr. Yen-Chien Kuo, Dr. Shang-Hsien Hsieh, and Dr. Chia-Hao Chen at NSRRC and Mr. Kui-Hon Ou Yang at National Taiwan University.

**Author contributions**

T.H.Y. and Y.W.L. conceived the idea and devised the project. H.C.H. and Y.C.C. conducted the material growth. T.H.Y., H.C.H., W.H.C., H.C.L., and Y.C.C. prepared samples for material characterizations. T.H.Y., H.C.H., and Y.F.C. performed the Raman, PL, and DOP analyses. T.H.Y. and H.C.H. conducted the SPEM investigation. F.X.C. and Y.H.K. performed the STM/S analysis. H.C.H. and A.C.C. conducted the AFM analyses. P.Y.L. conducted the HRTEM and SADP investigation. T.H.Y., H.C.L., B.W.L. and W.H.C. designed and fabricated devices, and performed the electrical measurement. T.H.Y. and Y.S.K. did the material simulation. The manuscript was written by T.H.Y., H.C.H., and F.X.C. with detailed discussion with all authors, and was revised by T.H.L., C.L.L. and Y.W.L. The overall project was supervised by T.H.L., C.L.L. and Y.W.L.

**Data availability**

The data that support the findings in this study are available from the corresponding authors upon reasonable request.

**Additional information**

Supplementary information is available in this paper.

**Competing interests**

The authors declare no competing interests.